\documentclass[runningheads]{llncs}
\usepackage[T1]{fontenc}
\usepackage{booktabs}
\usepackage{xcolor}
\usepackage{colortbl}
\usepackage{graphicx}
\usepackage{multirow}
\usepackage{hyperref} 
\hypersetup{
    colorlinks=true,
    citecolor=blue
}
\usepackage{amsmath}

\begin{document}
\title{SelfReg-UNet: Self-Regularized UNet for Medical Image Segmentation}
%
%
\def\thefootnote{*}\footnotetext{These authors contributed equally to this paper.}
\author{Wenhui Zhu\inst{1*} \and
Xiwen Chen\inst{2*} \and
Peijie Qiu\inst{3*}  \and
Mohammad Farazi\inst{1} \and
Aristeidis Sotiras\inst{3} \and Abolfazl Razi\inst{2} \and Yalin Wang\inst{1}
}
\authorrunning{Zhu. Wenhui et al.}
%
\institute{
School of Computing and Augmented Intelligence, Arizona State University, AZ, USA \and
School of Computing, Clemson University, SC, USA \and
McKeley School of Engineering, Washington University in St. Louis, MO, USA \\
}


\maketitle              
\begin{abstract}
Since its introduction, UNet has been leading a variety of medical image segmentation tasks. Although numerous follow-up studies have also been dedicated to improving the performance of standard UNet, few have conducted in-depth analyses of the underlying interest pattern of UNet in medical image segmentation. In this paper, we explore the patterns learned in a UNet and observe two important factors that potentially affect its performance: (i) irrelative feature learned caused by asymmetric supervision; (ii) feature redundancy in the feature map. To this end, we propose to balance the supervision between encoder and decoder and reduce the redundant information in the UNet. Specifically, we use the feature map that contains the most semantic information (i.e., the last layer of the decoder) to provide additional supervision to other blocks to provide additional supervision and reduce feature redundancy by leveraging feature distillation. The proposed method can be easily integrated into existing UNet architecture in a plug-and-play fashion with negligible computational cost. The experimental results suggest that the proposed method consistently improves the performance of standard UNets on four medical image segmentation datasets. The code is available at \url{https://github.com/ChongQingNoSubway/SelfReg-UNet}

\end{abstract}
\section{Introduction}
Medical image segmentation is a pivotal application for computer-aided diagnosis and image-guided systems. Recently, deep learning has ascended as the leading method in this domain, primarily attributed to the landmark contribution of UNet~\cite{Unet}. UNet defines a generic segmentation network architecture by leveraging the encoder to project semantic information into low-level features and the decoder to progressively upsample semantic features to segmentation masks. Many of its follow-up works~\cite{deeplabv3,MRUnet} have expanded this idea within the context of convolutional neural networks (CNN). Recently, numerous studies~\cite{transUnet,swimUnet,letvit,medT} have introduced vision transformer (ViT)~\cite{vit} to address the limitations of CNN by using a self-attention mechanism. Although ViT has a large receptive field and captures long-range dependencies between different image patches, it struggles to preserve fine-grained local context due to the lack of locality. 
To mitigate this problem, methods~\cite{hiformer,pvt-cascade} that bridge the gap between CNNs and ViTs (i.e., hybrid models) have been introduced in UNet design.
Note that these methods also bring much computational complexity and the number of parameters. Over-parameterization is a common issue in deep learning, often leading to feature redundancy and poor feature representation~\cite{featureredundancy,featureredundancy2,self-distillation}. However, this issue has not been formally investigated or considered in current medical segmentation models.

Beyond the aforementioned methods, some efforts focus on optimizing the architectural structures of the Unet. Following this vein, Att-Unet~\cite{atttentionUnet} proposes an attention-based skip connection to suppress irrelevant features. Unet++~\cite{Unet++} replaces the standard skip connections (i.e., concatenation/addition) with the nested dense skip pathways. UCTransNet~\cite{uctransUnet} thoroughly analyzes the effect of different skip connections and proposes a channel transformer to replace the conventional skip connection. These methods only investigate the information flow from encoder to decoder by manipulating the skip-connection. However, none of them have explored how to inform the encoder effectively via the learned features in the decoder, while our investigation reveals this information flow deserves more attention in a UNet.
This is mainly because the decoder receives more supervision than the encoder, which provides a natural way to filter out irrelevant information. 


In this paper, we perform empirical studies in two representative UNets (i.e., standard Unet~\cite{Unet} and SwinUnet~\cite{swimUnet}). Our analyses reveal two key findings: (i) Redundant features exist in the feature channel, with the shallow channels exhibiting more diversity than deep channels in a feature map; (ii) Asymmetric supervision between the encoder and the decoder in a UNet leads to semantic loss. This phenomenon diverges from observed trends in other computer vision tasks~\cite {fact,fact2}, where deep features exhibit to be more discriminative and can better localize the target of interest. 
To mitigate those issues, we introduce semantic consistency regularization and internal feature distillation to address semantic loss from asymmetric supervision and feature redundancy, respectively. This involves using more accurate semantics to supervise the other blocks and distilling information from shallow to deep channels in the feature map. 


\noindent \textbf{Contributions}: \textbf{(i)} Our exploration uncover asymmetric supervision and feature redundancy in UNet, suggesting a novel direction for future model design.
\textbf{(ii)} We suggest an orthogonal way to help the UNet discard irrelevant information and better preserve semantics by proposing a symmetric supervision regularization mechanism and leveraging the feature distillation.
\textbf{(iii)} The proposed methods can be seamlessly integrated into existing UNet frameworks (e.g., CNN-UNet and ViT-UNet), offering performance gains with minimal extra cost.

\section{Method}

\begin{figure}[!t]
\vspace{-1cm}
\centering
\includegraphics[width=0.95\columnwidth]{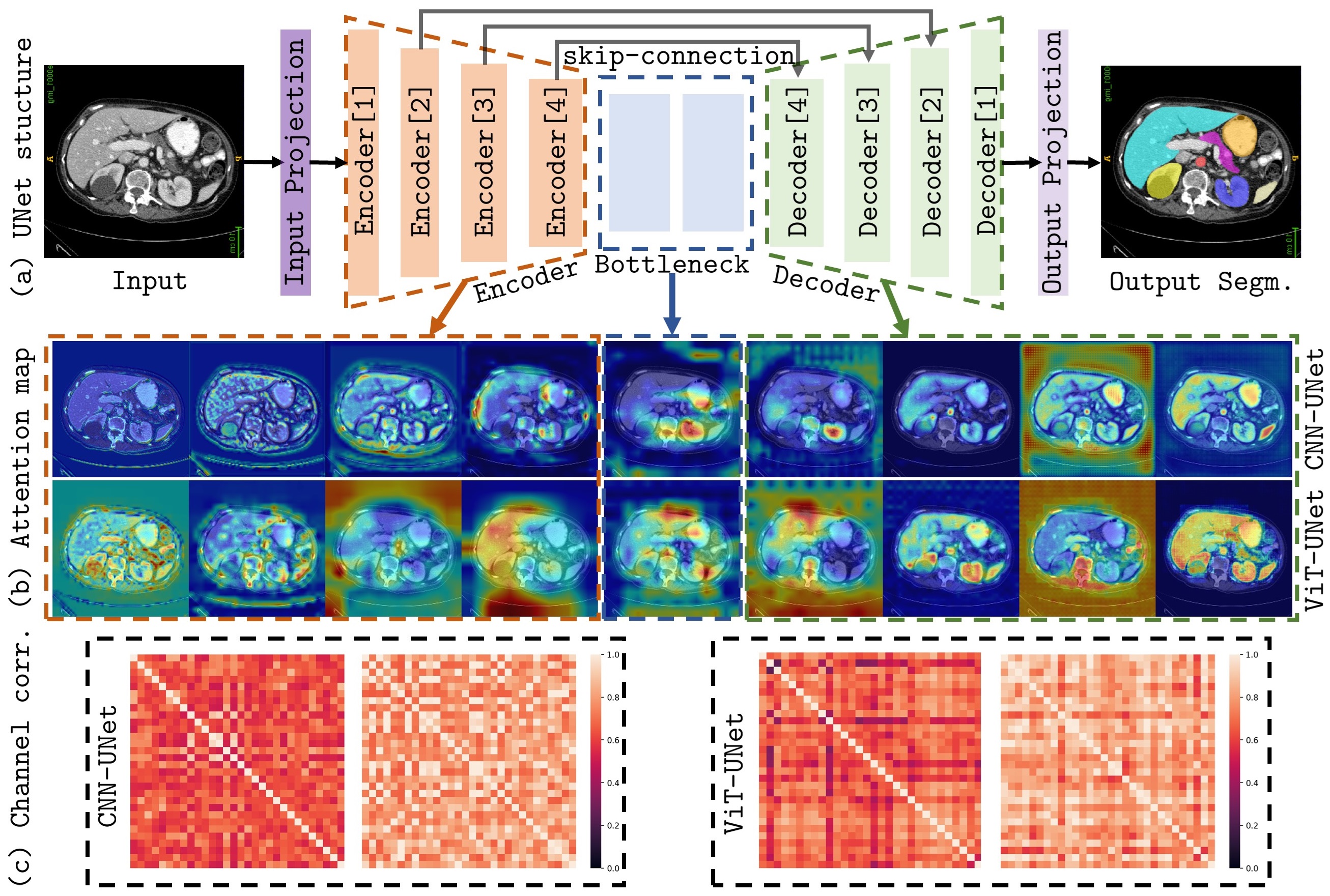} 
\caption{(a)Unet structure. (b) The attention map in Vit/CNN-based UNet corresponds to each encoder and decoder. (For more examples, refer to supplementary Appendix A) (c) ViT/CNN-based Unet feature similarity matrix between shallow (\texttt{Left}) and deeper channel (\texttt{Right}).}
\label{fig:network}
\end{figure}

\vspace{-0.1cm}
\subsection{Preliminary}
\vspace{-0.1cm}
In this paper, we take a standard Unet defined in~\cite{Unet} with a depth of 5 as an example. Without loss of generality, we consider a unified Unet structure for both CNN-based Unet~\cite{Unet} and ViT-based Unet~\cite{swimUnet} in our investigation, which consists of an input and output projection block, 4 encoder and decoder blocks, as well as a bottleneck (see Fig.~\ref{fig:network}(a)). For this purpose, we first define the patch embedding block in a ViT-Unet as the combination of the input projection block and the first encoder block. We then define the last up-sampling block in a ViT-UNet (i.e., last patch expanding in SwinUnet~\cite{swimUnet}) as the last decoder block. Each encoder/decoder block of the defined UNet comprises $L=2$ consective convolution/transformer blocks. The output projection block is a convolutional layer that maps the last feature map to a segmentation mask. 
For notation, we use \texttt{E}$_m^{(l)}$/ \texttt{D}$_{m}^{(l)}$ and \texttt{B}$^{(l)}$ to denote the $l$-th layer of the $m$-th encoder/decoder block and bottleneck, respectively. 
Accordingly, the corresponding feature map is denoted as $F_{m}^{l}$ following order from encoder to decoder (\texttt{E}$_1^{(1)}$, \texttt{E}$_1^{(2)}$,.\texttt{B}$^{(1)}$..,\texttt{D}$_2^{(2)}$,\texttt{D}$_1^{(1)}$).


\subsection{Analysis on features learned in a Unet}
We conduct analyses on features learned in UNets by employing two commonly used techniques: (i) gradient-weighted class activation mapping (Grad-CAM)~\cite{grad-cam}; (ii)  similarity analysis in a feature map.

\noindent \textbf{Asymmetric supervision in UNet.} 
We observed two interesting phenomena as evident in Fig.~\ref{fig:network}(b): (i) The learning patterns exhibit an asymmetry between the encoder and the decoder. The decoder could approximately locate some ground truth segmentation regions, and the encoder tends to capture unrelated information (\texttt{E}$_3$, \texttt{E}$_4$), dispersing the interest of patterns towards the boundaries.
(ii) In the decoder, the block (\texttt{D}$_1$) located further to the end demonstrates an accurate understanding of ground truth segmentation. Meahwhile, the blocks (\texttt{D}$_2$,\texttt{D}$_4$) learned the irrelevant information. The main reason lies in the different intensity of supervision signals each block receives. When tracing back from the output projection directly supervised by the ground truth, the supervision signal progressively diminishes. It leads to semantic loss, with some blocks (e.g., \texttt{E}$_1$,\texttt{E}$_3$,\texttt{E}$_4$,\texttt{B}) in the encoder even activating regions unrelated to segmentation. 



\noindent\textbf{Redundant Features in UNet.} Empirical investigations have shown that overparameterized CNN/ViT models tend to learn redundant features, leading to poor visual concepts~\cite{featureredundancy,featureredundancy2,self-distillation}. Taking the output of \texttt{E}$_1$ as an illustration, we calculate the feature similarity matrices in the channel dimension at both the shallow and the deep levels. As depicted in Fig.~\ref{fig:network}(c), we observe two phenomena in both ViT/CNN-based UNets: (i) Feature redundancy is prevalent in deep layers, with a high similarity matrix indicating the learning of similar features across channels. (ii) The shallow layers exhibit significant diversity, evidenced by a low similarity matrix. 
The over-parameterization that exists in Unet, akin to that in other networks, underlies these phenomena. The resultant redundant features often come with task-irrelevant visual features, leading to performance degradation and unnecessary computation overhead.

\subsection{Solutions}
\begin{figure}[!t]
\centering
\includegraphics[width=0.95\columnwidth]{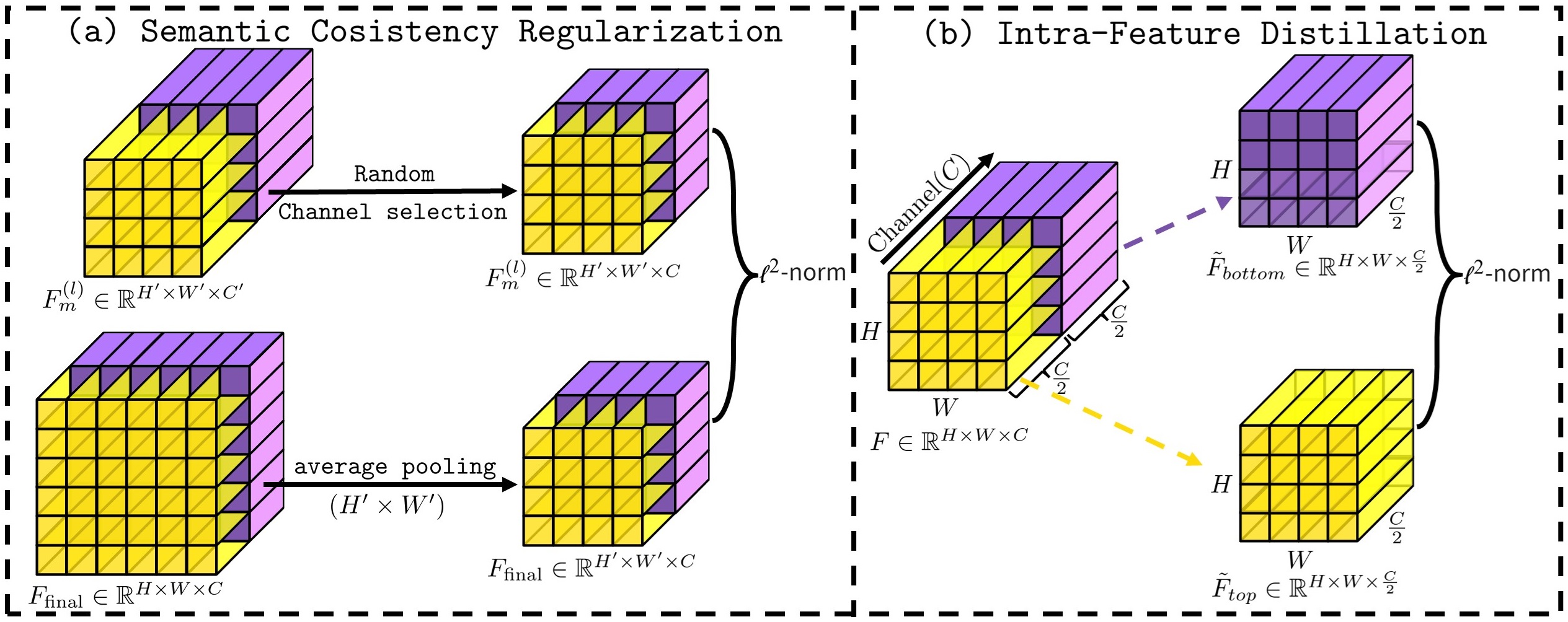} 
\caption{Demostrating the operation based on feature for (a) semantic consistency regularization and  (b) internal feature distillation.}
\vspace{-0.5cm}
\label{fig:method}

\end{figure}
\noindent\textbf{Semantic Consistency Regularization.}
Various studies have been proposed to tackle the loss of semantics in natural images by employing strategies such as knowledge distillation and feature alignment, with the goal of leveraging accurate features to guide those less informative ones~\cite{crosslayer,self-distillation,alignment,alignment2,alignment3}. Inspired by these works in natural images, we propose to use the feature map that contains the most semantic information (i.e., \texttt{D}$_1$ as observed) to provide additional supervision to the rest blocks in a UNet. As a result, we propose a generic paradigm, termed semantic consistency regularization (SCR), to balance the supervision between the encoder and the decoder. 
For simplicity, we demonstrate our idea using the feature distillation mechanism in~\cite{crosslayer,self-distillation} due to its popularity and simplicity (given as a mean square loss); while other knowledge distillation methods can be alternatives (given as KL divergence). To this end, we define the proposed SCR as 


\begin{equation}
    \mathcal{L}_{SCR} = \frac{1}{|M-1| |\mathcal{I}|} \sum_{m=1}^{M-1} \sum_{I \sim\mathcal{I}} \| \operatorname{RCS}(F_{m}^{i}) - \operatorname{AvgPool}(F_{final})\|^2, 
\end{equation}
Where $F_{final}$ is the feature map located at the last decoder block (\texttt{D}$_1$), and $F_{i}^{m}$ is all the feature map situated in the ith layer of the mth block (\texttt{E}$_1^{(1)}$, \texttt{E}$_1^{(2)}$,..., \texttt{D}$_3^{(2)}$,\texttt{D}$_2^{(1)}$,\texttt{D}$_2^{(2)}$) except the \texttt{D}$_1$. To align features in channel and spatial dimensional, we employed an average-pooling and random channel selection operation (RSC) as shown in Fig~\ref{fig:method}.(a). It is worth noting that the channel selection does not introduce extra modules~\cite{alignment,alignment2,alignment3}, reducing computation and semantic conflicts. The $L_{2}$ norm is used as a distance metric.



\noindent\textbf{Internal Feature Distillation.}
To solve the feature redundancy problem, some channel shrinkage methods have been proposed related to fields of model filter pruning, which leverage the $L_{p}$ norm penalty to induce a sparsity prior on channel salience~\cite{channel,channel2,channel3}. Inspired by this, we employed the $L_{p}$ norm for information distillation from shallow (top-half channel feature) to deep (bottom-half channel one), which guides the deeper features learned the useful context information. It can be formulated as: 

\begin{equation}
    \mathcal{L}_{IFD} = \frac{1}{|M| |\mathcal{I}|} \sum_{m=1}^{M} \sum_{I \sim\mathcal{I}} \| \widetilde{F_{m}^{i}} - \overline{F_{m}^{i}} \|^p
\end{equation}
where $F_{i}^{k}$ denotes all feature maps situated in the ith layer of the mth block (\texttt{E}$_1^{(1)}$, \texttt{E}$_1^{(2)}$,..., \texttt{D}$_2^{(1)}$,\texttt{D}$_2^{(2)}$,\texttt{D}$_1^{(1)}$), $\widetilde{F}$ is the deep channel feature, and $\overline{F}$ was shallow channel feature. As shown in Fig~\ref{fig:method}.(b), we partitioned the channels into top and bottom halves, using this division as boundaries to ensure the same number of features in both shallow and deep. Following~\cite{self-distillation,channel2,channel}, we employed the $L_{2}$ norm. Compared with methods that introduced extra modules for reducing redundancy~\cite{Unetredundant,Unetredundant2}, the $\mathcal{L}_{IFD}$ is simple and cost-free.

\noindent\textbf{Objective Function.}
The total loss is weighted of $\mathcal{L}_{SCR}$ and $\mathcal{L}_{IFD}$ with the standard combination of cross-entropy and dice losses $\mathcal{L}_{cd}$ ~\cite{uctransUnet,hiformer,swimUnet,Unet} that is evaluated between the prediction and the ground truth segmentation results. 
\begin{equation}
    \mathcal{L} = \mathcal{L}_{cd} + \lambda_{1} \mathcal{L}_{SCR} + \lambda_{2}\mathcal{L}_{IFD}, 
\end{equation}
where $\lambda_{1}$ and $\lambda_{2}$ are balance parameters.

\section{Experiments and Results}
\subsection{Dataset}

\noindent\textbf{Synapse Multi-Organ Segmentation.} 
Following~\cite{swimUnet,transUnet,hiformer}, the Synapse dataset comprises 30 cases comprising 3779 axial abdominal clinical CT images. Of these, 18 samples are designated for training, while 12 are reserved for testing. We evaluate performance on eight abdominal organs using the average Dice Similarity Coefficient (DSC) as the evaluation metric.

\noindent\textbf{Automated cardiac diagnosis challenge dataset.}
The ACDC dataset comprises 100 cardiac MRI scans from diverse patients, each labeled with left ventricle (LV), right ventricle (RV), and myocardium (MYO). Following~\cite{transUnet,pvt-cascade}, we allocate 70 cases (1930 axial slices) for training, 10 for validation, and 20 for testing. We evaluate our method using the DSC as the evaluation metric.

\noindent\textbf{Nuclear segmentation and Gland segmentation.} The Gland segmentation dataset (GlaS)~\cite{GlaS} has 85 images for training and 80 for testing. The Multi-Organ Nucleus Segmentation (MoNuSeg) dataset~\cite{MoNuSeg} has 30 images for training and 14 for testing. Following~\cite{uctransUnet}, we perform the three times 5-fold cross-validation results on GlaS and MoNuSeg datasets. The average DSC and IoU are used as evaluation metrics.

\begin{figure*}[hb]
\vspace{-0.4cm}
\centering
\includegraphics[width=1\columnwidth]{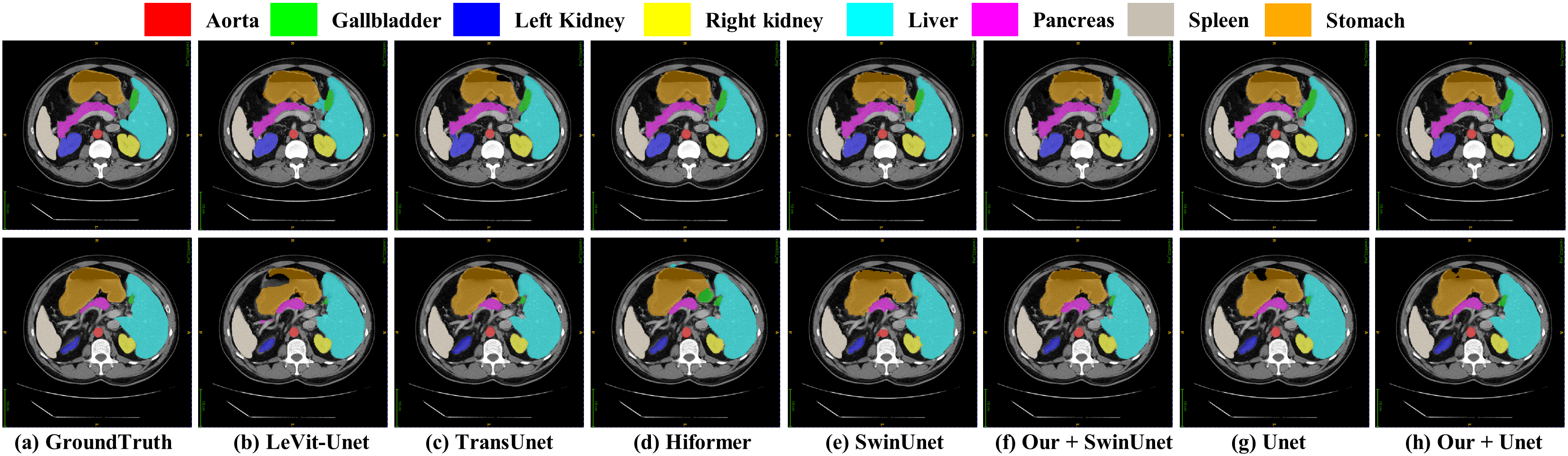} 
\caption{Comparison of segmentation performance in Synapse dataset. }
\label{fig:seg1}
\vspace{-1cm}
\end{figure*}

\begin{figure*}[b]
\vspace{-0.3cm}
\centering
\includegraphics[width=1\columnwidth]{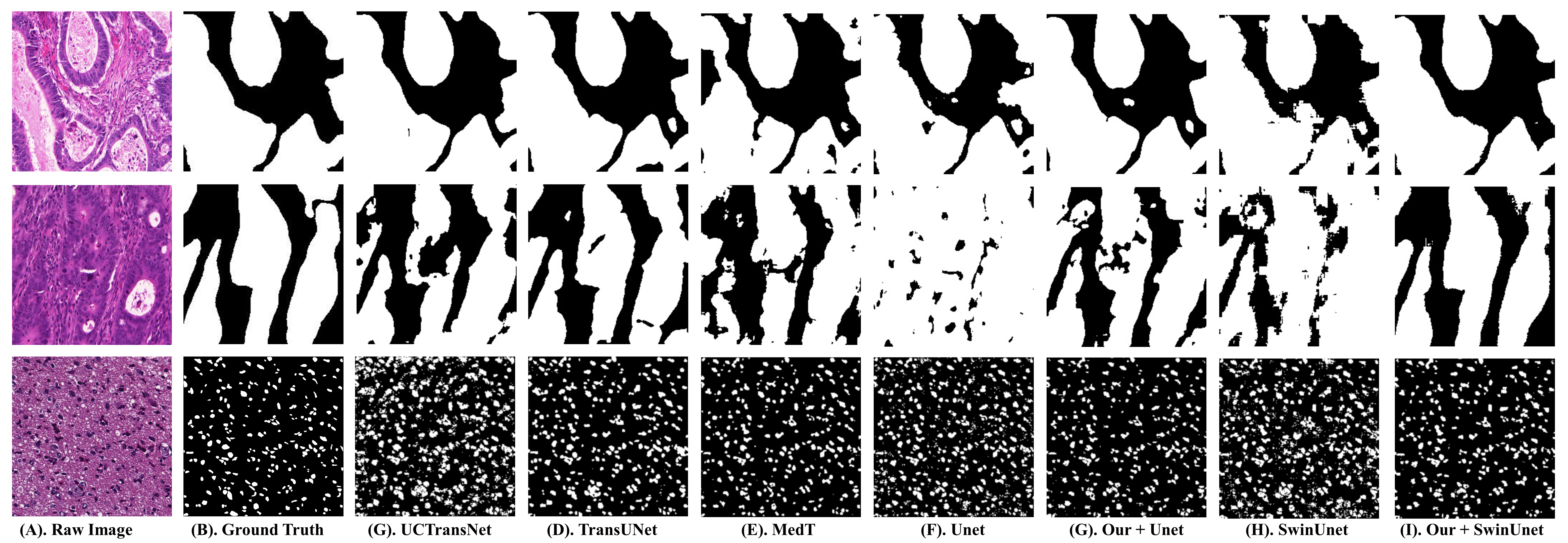} 
\caption{Comparison of segmentation performance in Glas and MoNuSeg dataset. }
\label{fig:seg2}
\end{figure*}


\subsection{Experiment settings}
We evaluate the effectiveness of our proposed loss on both SwinUnet and Unet in these four datasets, and the training setups (i.e., batch size, optimizer, learning rate, etc.) are consistent with \cite{swimUnet}. All experiments were conducted with an input image size of 224 x 224 and the same data augmentation and preprocessing in~\cite{hiformer,pvt-cascade,uctransUnet}, using an Nvidia GTX3090 with 24GB of memory for training. Following in~\cite{swimUnet,Unet}, pre-trained weights from ImageNet were employed in SwinUnet, while Unet was trained from scratch.

\noindent\textbf{Comparison with state-of-the-art methods.} We compare our methods with recent SOTA models, including R50 Unet~\cite{transUnet}, Att-Unet~\cite{atttentionUnet}, Unet++~\cite{Unet++}, TransUnet~\cite{transUnet},swinUnet~\cite{swimUnet}, Levit-Unet~\cite{letvit}, DeepLabv3~\cite{deeplabv3}, HiFormer~\cite{hiformer}, PVT-cascade~\cite{pvt-cascade}, UCTransNet~\cite{uctransUnet}, MedT~\cite{medT}. 


\begin{table*}[t]
\centering
\caption{Comparison with SOTA methods on Synapse multi-organ CT dataset. $\Delta$ denotes the improvement gain (\%) by comparing with the original method.}
\label{tab:synapse}
\resizebox{\textwidth}{!}{%
\begin{tabular}{l|c|cccccccc}
\toprule
\textbf{Methods} & \textbf{Average DSC} &  \textbf{Aorta} & \textbf{Gallbladder} & \textbf{Kidney(L)} & \textbf{Kidney(R)} & \textbf{Liver} & \textbf{Pancreas} & \textbf{Spleen} & \textbf{Stomach} \\ \hline
R50 U-Net & 74.68 & 87.74 & 63.66 & 80.60 & 78.19 & 93.74 & 56.90 & 85.87 & 74.16 \\
U-Net & 76.85 & 89.07 & 69.72 & 77.77 & 68.60 & 93.43 & 53.98 & 86.67 & 75.58 \\
R50 Att-Unet  & 75.57 & 55.92 & 63.91 & 79.20 & 72.71 & 93.56 & 49.37 & 87.19 & 74.95 \\
Att-Unet & 77.77 & 89.55 & 68.88 & 77.98 & 71.11 & 93.57 & 58.04 & 87.30 & 75.75 \\
TransUnet  & 77.48 & 87.23 & 63.13 & 81.87 & 77.02 & 94.08 & 55.86 & 85.08 & 75.62 \\
SwinUnet  & 79.13  & 85.47 & 66.53 & 83.28 & 79.61 & 94.29 & 56.58 & 90.66 & 76.60 \\
LeVit-Unet & 78.53  & 78.53 & 62.23 & 84.61 & 80.25 & 93.11 & 59.07 & 88.86 & 72.76 \\
DeepLabv3 & 77.63  & 88.04 & 66.51 & 82.76 & 74.21 & 91.23 & 58.32 & 87.43 & 73.53 \\
HiFormer  & 80.29 & 85.63 & 73.29 & 82.39 & 64.84 & 94.22 & 60.84 & 91.03 & 78.07 \\
\hline
\rowcolor{blue!8}
Our + Unet & \underline{80.34} & 88.74 & 71.78 & 85.32 & 80.71 & 93.80 & 62.22 & 84.78 & 75.39  \\ 
\rowcolor{blue!8}

$\Delta$   &$\textcolor{red}{+3.49}$ & \textcolor{blue}{-0.33} & \textcolor{red}{$+2.06$} & \textcolor{red}{$+7.55$} & \textcolor{red}{$+12.11$} & \textcolor{red}{$+0.37$} & \textcolor{red}{$+8.24$} & \textcolor{blue}{$-1.89$} & \textcolor{blue}{$-0.19$} \\
\hline
\rowcolor{blue!8}
Our + SwinUnet & \textbf{80.54} & 86.07 & 69.65 & 85.12 & 82.58 & 94.18 & 61.08 & 87.42 & 78.22 \\
\rowcolor{blue!8}
$\Delta$   & \textcolor{red}{$+ 1.41$} & \textcolor{red}{$+0.60$} & \textcolor{red}{$+3.12$} &  \textcolor{red}{$+$1.84} & \textcolor{red}{$+2.97$} &  \textcolor{blue}{$-0.11$} & \textcolor{red}{$+4.50$} & \textcolor{blue}{$-3.24$} &  \textcolor{red}{$+1.62$}\\
\hline
\end{tabular}%
}
\end{table*}

\begin{table}[!t]
\vspace{-0.2cm}
    \centering
    \begin{minipage}{0.4\columnwidth}
        
        \centering
        \caption{Comparison of different methods in ACDC dataset.}
        \label{tab:addc}
        \resizebox{1\columnwidth}{!}{
       \begin{tabular}{l|c|c|c|c} 

\hline
\textbf{Methods} & \textbf{Avg DSC} & \textbf{RV} & \textbf{Myo} & \textbf{LV} \\
\hline
R50 + AttnUnet & 86.75 & 87.58 & 79.2 & 93.47 \\
ViT + CUP & 81.45 & 81.46 & 70.71 & 92.18 \\
Unet & 89.68 & 87.17 & 87.21 & 94.68 \\
TransUnet & 89.71 & 86.67 & 87.27 & 95.18 \\
SwinUnet  & 88.07 & 85.77 & 84.42 & 94.03 \\
LeVit-Unet & 88.21 & 85.56 & 84.75 & 94.32\\
Hiformer & 90.82 & 88.55 & 88.44 & 95.47\\
PVT - CASCADE  & 90.45 & 87.20 & 88.96 & 95.19 \\
\hline
\rowcolor{blue!8}
Our + Unet  & \underline{91.43} & 88.92 & 89.49 & 95.88\\
\rowcolor{blue!8}
$\Delta$  & \textcolor{red}{$+1.75$} & \textcolor{red}{$+1.75$} & \textcolor{red}{$+2.28$}& \textcolor{red}{$+1.20$}\\ 
\hline
\rowcolor{blue!8}
 Our + SwinUnet & \textbf{91.49} & 89.49 & 89.27 & 95.70 \\ 
\rowcolor{blue!8}
$\Delta$  & \textcolor{red}{$+3.42$} & \textcolor{red}{$+3.72$} & \textcolor{red}{$+4.85$}& \textcolor{red}{$+1.67$}\\ 
\hline
\end{tabular}%
}
    \end{minipage}%
    \hfill
    \begin{minipage}{0.52\columnwidth}
    \caption{Comparison of different methods in Glas and MoNuSeg datasets.}
    \label{tab:glas}
        \centering
        \resizebox{1\columnwidth}{!}{
        \begin{tabular}{l|cc|cc}
\hline
\multicolumn{1}{c}{\multirow{2}{*}{\textbf{Method}}} & \multicolumn{2}{c}{\textbf{Glas}} & \multicolumn{2}{c}{\textbf{MoNuSeg}} \\
\multicolumn{1}{c}{} & \multicolumn{1}{c}{ \textbf{DSC (\%)}} &\multicolumn{1}{c}{ \textbf{IOU (\%)}}& \multicolumn{1}{c}{ \textbf{DSC (\%)}} &\multicolumn{1}{c}{\textbf{ IOU (\%)}}  \\
\hline
U-Net  & 85.45$\pm$1.25 & 74.78$\pm$1.67 & 76.45$\pm$2.62 & 62.86$\pm$3.00 \\
Unet++  & 87.56$\pm$1.17 & 79.13$\pm$1.70 & 77.01$\pm$2.10 & 63.04$\pm$2.54 \\
AttUnet & 88.80$\pm$1.07 & 80.69$\pm$1.66 & 76.67$\pm$1.06 & 63.47$\pm$1.16 \\
MRUnet & 88.73$\pm$1.17 & 80.89$\pm$1.67 & 78.22$\pm$2.47 & 64.83$\pm$2.87 \\
TransUnet & 88.40$\pm$0.74 & 80.40$\pm$1.04 & 78.53$\pm$1.06 & 65.05$\pm$1.28 \\
MedT & 85.92$\pm$2.93 & 75.47$\pm$3.46 & 77.46$\pm$2.38 & 63.37$\pm$3.11 \\
SwimUnet & 89.58$\pm$0.57 & 82.06$\pm$0.73 & 77.69$\pm$0.94 & 63.77$\pm$1.15 \\
UCTransNet  & 90.18$\pm$0.71  & 82.96$\pm$1.06  & 79.08$\pm$0.67 & 65.50$\pm$0.91 \\
\hline
\rowcolor{blue!8}
Our + Unet & \underline{90.93$\pm$0.20}  & \underline{84.08$\pm$0.41}  & \textbf{80.18$\pm$0.19} & \textbf{67.07$\pm$0.29} \\
\rowcolor{blue!8}
$\Delta$  & \textcolor{red}{$+5.48$} & \textcolor{red}{$+9.3$} & \textcolor{red}{$+3.73$}& \textcolor{red}{$+4.21$}\\ 
\hline
\rowcolor{blue!8}
Our + SwinUnet & \textbf{91.62$\pm$0.16}  & \textbf{85.29$\pm$0.30}  & \underline{79.38$\pm$0.15} & \underline{65.87$\pm$0.21} \\
\rowcolor{blue!8}
$\Delta$  & \textcolor{red}{$+2.04$} & \textcolor{red}{$+3.23$} & \textcolor{red}{$+1.69$}& \textcolor{red}{$+2.10$}\\

\hline

\end{tabular}
}
        
    \end{minipage}
\vspace{-0.3cm}
\end{table}


\subsection{Results}

The results are shown in Table~\ref{tab:synapse} for Synapse dataset, Table~\ref{tab:addc} for ACDC dataset, and Table~\ref{tab:glas} for Glas and MoNuSeg datasets. The main observation is our proposed loss is effective and can lead to substantial gains. Specifically, by utilizing our proposed loss, Unet obtains a 3.49\%, 1.75\%, 5.48\%, and 3.73\% improvement in average DSC in the four datasets, respectively. Likewise, in these four datasets, SwinUnet can obtain a 1.41\%, 3.42\%, 2.04\%, and 1.69\% DSC improvement. 

Consequently, our enhanced models can outperform the previous State-of-the-art (SOTA) methods. For example, our enhanced Unet and SwinUnet exhibit improvements of 0.05\% and 0.25\% in average DSC, respectively. Compared with Hiformer, TransUnet, Unet, and swinUnet~\cite{Unet,swimUnet,hiformer,transUnet}, the performance gains of our enhanced models primarily derive from tackling challenging organ segmentation tasks such as the gallbladder, left kidney, right kidney and pancreas, as illustrated in Fig~\ref{fig:seg1} and Table~\ref{tab:synapse}. Similarly, in ACDC dataset, compared with SOTA HiFormer~\cite{hiformer}, our Unet and swinUnet methods demonstrate enhancements of 0.6\% and 0.67\% in average DSC, respectively. This superiority is also generalized to Glas and MoNuSeg datasets, and our methods exhibit boosted performance over the previous SOTA method, UCTransNet, with 0.7\% and 1.1\% DSC gain in both datasets, respectively. Fig~\ref{fig:seg2} demonstrates our method's notable improvement in Glas dataset. Note that Unet with optimized skip connections (e.g., UCTTransNet~\cite{uctransUnet}) leads to irrelevant segmentation and incomplete shapes. Especially segments resembling the background (last row). While SwinUnet, with our proposed loss, demonstrates results that are closely comparable to the ground truth, featuring full shapes and clear backgrounds, particularly in the hard to improvement sample (the third row in Fig~\ref{fig:seg2}).
These observations support that our approach can provide contextual supervision, ensuring correct semantics across blocks and preventing the learning of irrelevant features.

 



\begin{figure}[t]
\centering
\includegraphics[width=1\columnwidth]{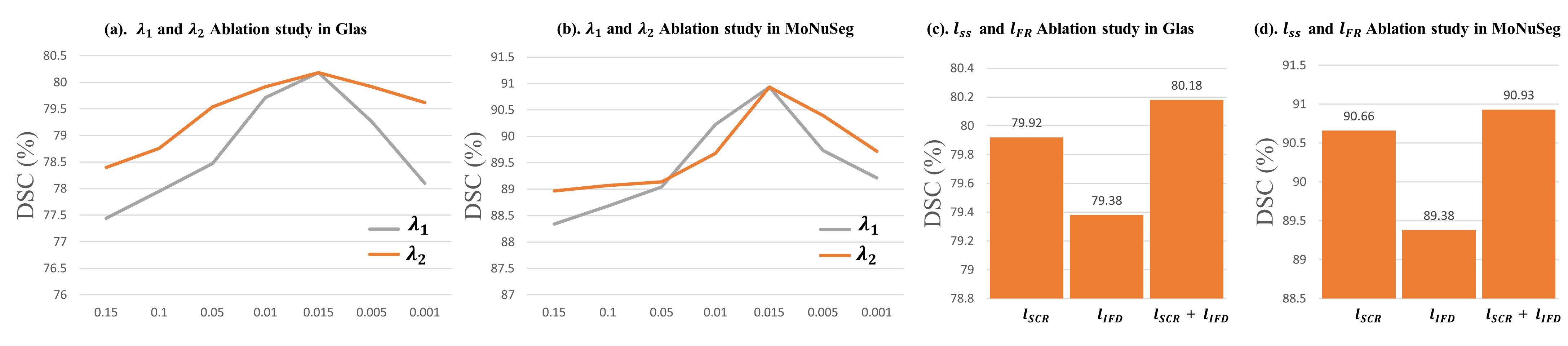} 
\caption{Ablation study for balance parameters $\lambda$ and loss based on Glas and MoNuSeg. }
\label{fig:ablation}
\vspace{-0.5cm}
\end{figure}

\subsection{Ablation studies}
To look deeper into our method, we perform a series
of ablation studies using Unet on the
Glas and MoNuSeg dataset. The results and analysis are as follows:

\noindent\textbf{Loss balance hyperparameters.} As result shown in Fig~\ref{fig:ablation}(a,b), we observe that the $\lambda_{1} = 0.015$ and $\lambda_{2} = 0.015$ is the optimal setting. As the weight of $\lambda_{1}$ decreases ($l_{SCR}$), the performance deteriorates rapidly. This also indicates that the lack of correct semantic supervision is a reason for the performance decline.

\noindent\textbf{Effectiveness of the proposed losses.} We conducted the ablation study with the different losses as shown in Fig~\ref{fig:ablation}(c,d). Both datasets exhibit consistent results, achieving optimal performance when both losses are used simultaneously.

\vspace{-0.2cm}
\section{Conclusion}
\vspace{-0.2cm}
In this paper, we unveil the problem of asymmetric supervision and feature redundancy in Unet-based medical image segmentation, suggesting a novel approach through loss optimization that incorporates semantic consistency regularization and internal feature distillation. Our experimental results demonstrate that addressing these two issues indeed has the potential to improve Vit/CNN-based UNet models, and the proposed method holds potential across a wide range of medical image segmentation tasks. 
In the future, our findings can aid in the design of UNet, and we will explore additional solutions from this perspective.
%
%
%
%
\bibliographystyle{splncs04}
\bibliography{refs}

\end{document}


\title{Supplementary Material For SelfReg-UNet}
%
%
%
\author{}

\institute{}

\maketitle              
\appendix
\vspace{-1cm}
\section{More Examples For Attention Map}

\begin{figure*}[h]
\vspace{-0.8cm}
\centering
\includegraphics[width=1\columnwidth]{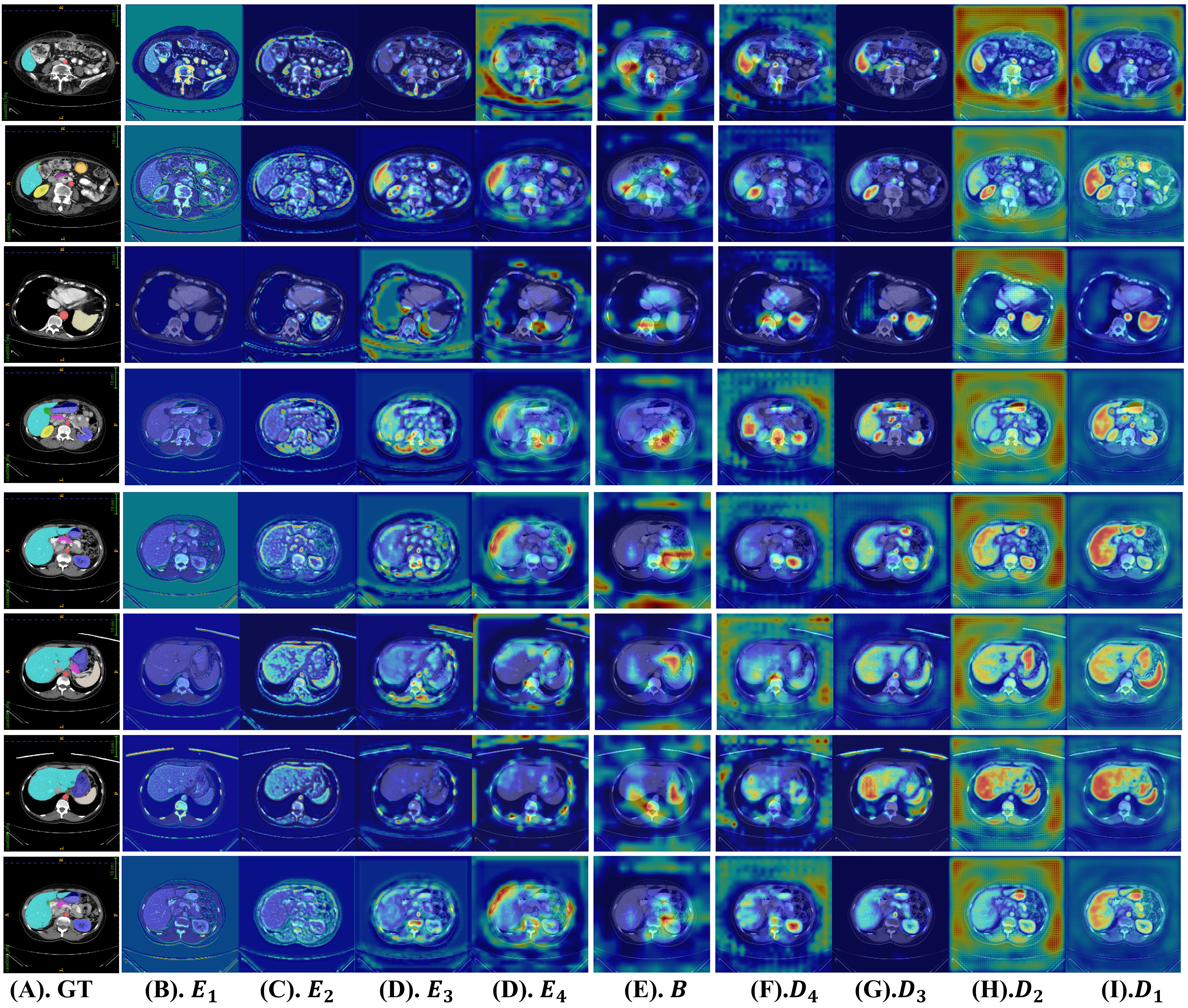}
    \caption{Attention map visualization of each block for vanilla/CNN UNet~\cite{unet}. GT: ground truth}
\label{fig:unet}
\vspace{-0.3cm}
\end{figure*}

\begin{figure*}[!h]
\vspace{-0.5cm}
\centering
\includegraphics[width=0.8\columnwidth]{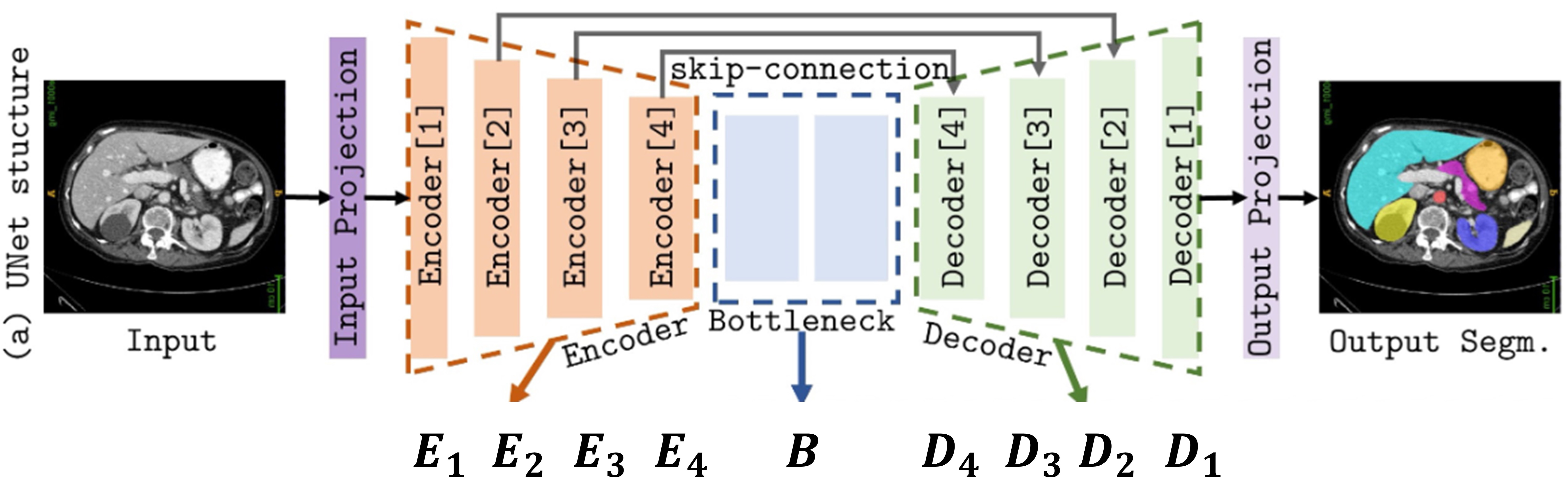}
    \caption{UNet structure and corresponding to each block of encoder/bottleneck/decoder. Consistent with the manuscript, each block includes two CNN/Transformer layers. Here, we have only visualized the last layer in each block.}
\label{fig:network}
\end{figure*} 

\begin{figure*}[!t]
\vspace{-1cm}
\centering
\includegraphics[width=1\columnwidth]{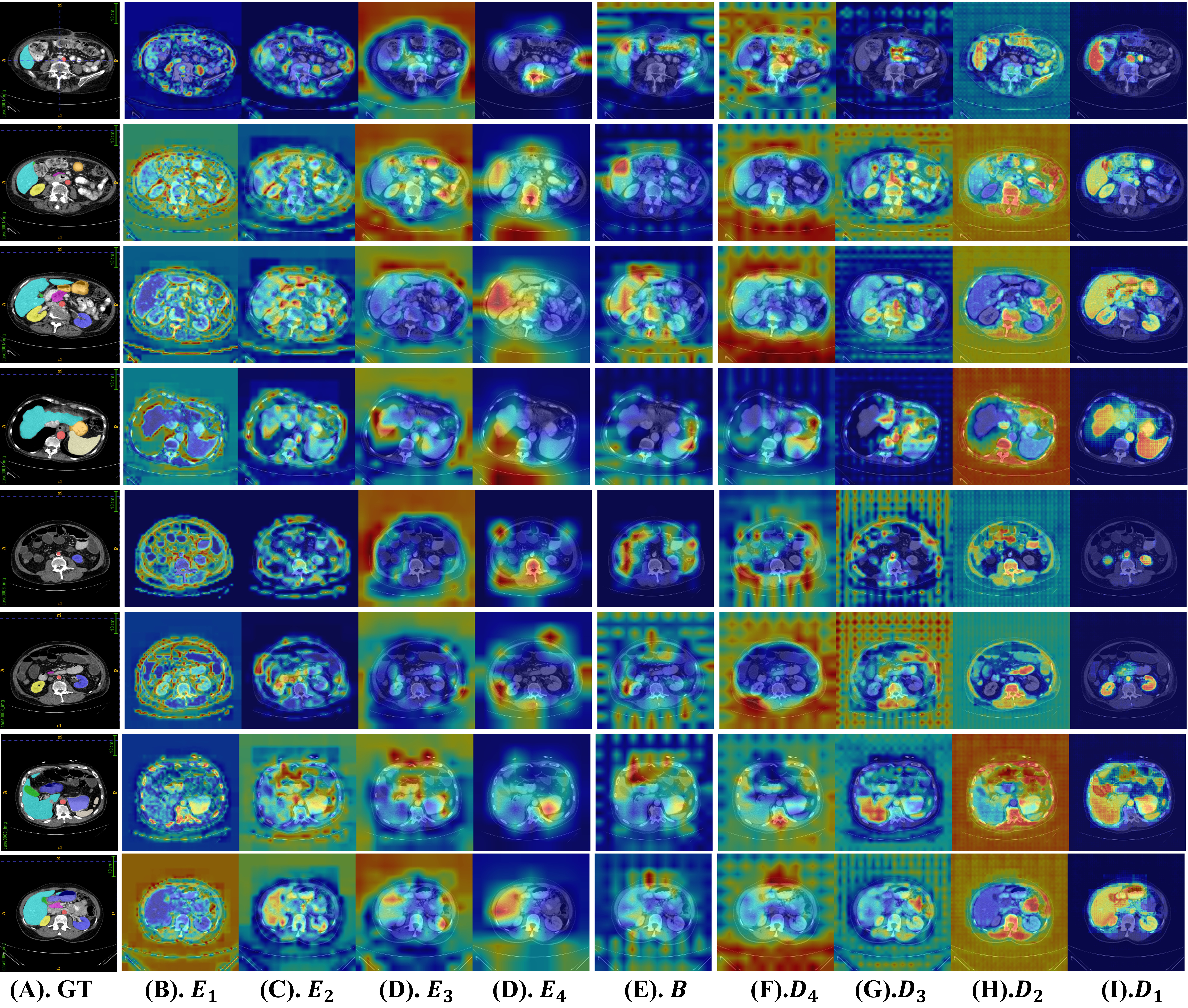}
    \caption{Attention map visualization of each block for Swin UNet~\cite{swimunet}.GT: ground truth }
\label{fig:swim}
\vspace{-0.5cm}
\end{figure*}

\small
\bibliographystyle{splncs04}
\bibliography{refs}